\documentclass[prb,a4paper,showpacs,superscriptaddress]{revtex4}
\usepackage{amsmath} 
\usepackage{amsfonts}
\usepackage{amssymb}
\usepackage{bm} 
\usepackage{graphicx}

\newcommand{\imagu}{{\rm i}}
\begin{document}
\title{Calculation of NMR Properties of Solitons in Superfluid $^3$He-A}
\author{R. H\"anninen}
\affiliation{Low Temperature Laboratory, Helsinki University of
Technology, P.O. Box 2200, FIN-02015 HUT, Finland } 
\author{E. V. Thuneberg}
\affiliation{Low Temperature Laboratory, Helsinki University of
Technology, P.O. Box 2200, FIN-02015 HUT, Finland } 
\affiliation{Department of Physical Sciences,
P.O.Box 3000, FIN-90014 University of Oulu, Finland}

\date{\today}

\begin{abstract}
Superfluid $^3$He-A has domain-wall-like structures, which are called
solitons. We calculate numerically the structure of a splay soliton. We
study the effect of solitons on the nuclear-magnetic-resonance
spectrum by calculating the frequency shifts and the amplitudes of the
soliton peaks for both longitudinal and transverse oscillations of
magnetization. The effect of dissipation caused by normal-superfluid
conversion and spin diffusion is calculated.  The calculations are in
good agreement with experiments, except a problem in the transverse
resonance frequency of the splay soliton or in magnetic-field
dependence of reduced resonance frequencies.
\end{abstract}

\maketitle

\section{INTRODUCTION}

Nuclear magnetic resonance (NMR) has turned out to be very useful
for studying the superfluid phases of liquid
$^3$He. The two superfluid phases A and B are distinguished in the
NMR spectrum by different frequency shifts of the absorption
peaks.\cite{Leggett74} In addition to these ``bulk'' peaks, one
often observes additional ``satellite'' peaks. These are caused by
topological objects and textures that appear in the superfluid order
parameter. Especially in superfluid $^3$He-A, several different objects
have been identified based on the frequency
shifts of the satellite peaks.\cite{vollhardtwolfle,Volovik,Lounas} The
simplest of these are solitons. They are domain-wall-like structures
where a planar object separates two different but degenerate bulk
states.

The satellite peaks in $^3$He-A were first observed in
measurements in the mid 1970's.\cite{Avenel,GouldLee} 
Soon after the theory of solitons in $^3$He was developed by Maki
and Kumar.\cite{MakiKumar77,MakiKumar} Their
calculation gave a striking agreement with the measured frequency
shifts of the satellite peaks at temperatures close to the superfluid
transition temperature $T_{\rm c}$.  This initial success had the
consequence that further studies of solitons went to other
directions\cite
{BruinsmaMaki,GouldBartolacBozler,KharadzeMaki,NakaharaMaki} and,
unfortunately, no more precise calculations were done.

A soliton has two basic structures, ``twist''
and ``splay,'' which correspond to the cases of a magnetic field ($B\gg
1$  mT) perpendicular and parallel, respectively, to the
plane of the soliton wall. Both these structures can be studied using
small oscillations of the magnetization that are either transverse or
longitudinal relative to the static field.

There are several points that can be improved in the previous
calculations, given as follows. (i) The structure of the splay soliton
was calculated only by using variational approximation. (ii) The
calculations were limited to temperatures near $T_{\rm c}$. (iii) The
effect of different parameter values was not studied. (iv) The
amplitudes and widths of the satellite peaks as well as peaks of
higher order were not studied. (v) Dissipation was neglected. 
It is just these points  that we address in this paper. An additional
motivation is that the study of solitons opens the way to detailed
understanding of more complicated topological objects such as vortices.

We find that, using dissipationless theory, the agreement between the
theoretical and experimental frequency shifts is generally very good.
However, we find a puzzling disagreement in the transverse oscillation
frequency of the splay soliton. This disagreement has remained unnoticed
because no detailed comparison between theory and experiment has been
published. Further, we find that the theory is particularly inflexible to
explain this discrepancy away. Taking into account dissipation, in
particular spin diffusion, changes these conclusions. On one hand,
the  disagreement in the splay-soliton frequency is reduced. On the
other hand, we find considerable extra shift of resonance frequencies in
high fields, which has not been reported experimentally. We also
point out that the longitudinal resonance of the splay soliton has not
been studied experimentally. Measurement of these quantities would be
important to test our understanding of the basic properties of
superfluid $^3$He.

We start in Sec.~\ref{s.hyrdo} with a short introduction to the 
hydrodynamic theory and NMR in $^3$He-A. In Sec.~\ref{s.eqstruc} we
solve numerically the structure of the splay soliton. The frequency and
the absorption of the principal satellite peak are determined in
Secs.~\ref{s.princnmr} and \ref{s.intes} ignoring dissipative 
effects. In Sec.~\ref{s.intes} we calculate the frequencies and
absorption of the higher modes. In Sec.~\ref{s.dissipation} we take
into account dissipation and calculate the effect of the spin diffusion
and normal-superfluid relaxation on the absorption spectrum.

\section{HYDRODYNAMIC THEORY}\label{s.hyrdo}

Here we briefly present some main points of the hydrodynamic theory and
NMR in $^3$He-A. The order parameter of superfluid $^3$He-A is a
$3\times 3$ tensor of the form\cite{Leggett,vollhardtwolfle}   
\begin{equation}
A_{\mu j}=\Delta\hat{d}_\mu(\hat{m}_j+{\rm i}\hat{n}_j) ,
\label{e.ordpar} 
\end{equation} 
where $\hat{\bf d}$, $\hat{\bf m}$ and $\hat{\bf n}$ are unit vectors
and $\hat{\bf m}\perp \hat{\bf n}$. It is conventional to define
$\hat{\bf l}=\hat{\bf m}\times\hat{\bf n}$, which gives the axis of
the orbital angular momentum of a Cooper pair. The unit vector
$\hat{\bf d}$ defines the axis along which the spin of the Cooper pair
vanishes. In a static magnetic field, the structure of a soliton can
be determined by finding a local minimum
for the free energy\cite{vollhardtwolfle,Cross,Fetterrev}   
\begin{equation}
F_{\rm static}=\int d^3r(f_{\rm d}+f_{\rm g}+f_{\rm h}).
\label{e.fstatic}
\end{equation}
Here $f_{\rm d}$ comes from the dipole-dipole interaction between
nuclear moments, 
\begin{equation}
f_{\rm d}=-\frac{1}{2}\lambda_{\rm d}(\hat{\bf d}\cdot\hat{\bf l})^2,
\label{e.fdip}
\end{equation}
$f_{\rm h}$ from coupling to the external field ${\bf B}$,
\begin{equation}
f_{\rm h}=\frac{1}{2}\lambda_{\rm h}(\hat{\bf d}\cdot{\bf B})^2\ ,
\label{e.fmag}
\end{equation}
and $f_{\rm g}$ from the gradient of the order parameter,
\begin{eqnarray}
2f_{\rm g} &=& \rho_\perp{\bf v}_s^2 
+ (\rho_\parallel-\rho_\perp)(\hat{\bf l}\cdot{\bf v}_s)^2 
+ 2C{\bf v}_s\cdot\mbox{\boldmath$\nabla$}\times\hat{\bf l} 
\nonumber \\ 
&-& 2C_0(\hat{\bf l}\cdot{\bf v}_s)(\hat{\bf
l}\cdot\mbox{\boldmath$\nabla$}\times\hat{\bf l})  
+ K_{\rm s}(\mbox{\boldmath$\nabla$}\cdot\hat{\bf l})^2 
+ K_{\rm t}(\hat{\bf l}\cdot\mbox{\boldmath$\nabla$}
\times\hat{\bf l})^2 \nonumber \\
&+& K_{\rm b}\vert\hat{\bf l}\times(\mbox{\boldmath$\nabla$}
\times\hat{\bf l})\vert^2 
+ K_5\vert(\hat{\bf l}\cdot\mbox{\boldmath$\nabla$})\hat{\bf d}\vert^2 
+ K_6 \sum_{i,j}[(\hat{\bf l}\times\mbox{\boldmath$\nabla$})_i
\hat{d}_j]^2. 
\label{e.fgrad}
\end{eqnarray} 
The gradient energy (\ref{e.fgrad}) also includes the 
kinetic energy arising from the superfluid velocity ${\bf v}_s =
(\hbar/2m_3) \sum_{i} \hat{m}_i
\mbox{\boldmath$\mbox{\boldmath$\nabla$}$}\hat{n}_i$, where $m_3$ is the
mass of a $^3$He atom. However, in the following we limit to the case
of zero superfluid velocity.  The parameters appearing in the gradient
energy have been calculated in the weak-coupling approximation by
Cross\cite{Cross} and D\"orfle,\cite{Dorfle} the latter including more
Fermi-liquid parameters. For numerical values see  Refs.\
\onlinecite{Hook} and \onlinecite{web}. The characteristic scales are
given by the dipole length $\xi_{\rm d} =
(\hbar/2m_3)\sqrt{\rho_\parallel/\lambda_{\rm d}} \approx 10\     
\mu{\rm m}$ and the dipole field $B_{\rm d} =\sqrt{\lambda_{\rm
d}/\lambda_{\rm h}} \approx 2\ {\rm mT}$. 

We consider a static field ${\bf B}_0 = B_0 \hat{\bf z}$. We assume
that the equilibrium $\hat{\bf d}$, denoted by $\hat{\bf d}_0$, lies
in the plane perpendicular to ${\bf B}_0$: 
\begin{eqnarray}
\hat{\bf d}_0 = \hat{\bf x}\cos\theta + \hat{\bf y}\sin\theta.
\label{e.d0}
\end{eqnarray}
This situation is always achieved in large field $B_0
\gg B_{\rm d}$, where $\hat{\bf d}$ is forced to the plane by $f_{\rm
h}$ (\ref{e.fmag}), but in some cases this happens in low fields as
well.  Minimization of the total energy (\ref{e.fstatic}) gives for
$\theta$ the equation 
\begin{equation}
\mathcal{D}\theta+(\hat{\bf l} \cdot \hat{\bf d}_0)
 (\hat{\bf l}\times\hat{\bf d}_0)_z=0, 
\label{e.dtheta}\end{equation} 
where the operator $\mathcal{D}$ is defined by
\begin{equation}
\mathcal{D}f=-\frac{K_6}{\lambda_{\rm d}}\nabla^2f
-\frac{K_5-K_6}{\lambda_{\rm d}}
\mbox{\boldmath$\nabla$}\cdot\left\lbrack\hat{\bf l}  (\hat{\bf l}\cdot
\mbox{\boldmath$\nabla$}) f\right\rbrack.
\label{e.ddef}
\end{equation} 

In a dynamic magnetic state one has to include the spin magnetization
$\gamma{\bf S}$ as a new variable in addition to $\hat{\bf d}$ and
$\hat{\bf l}$. The effective energy density has the
form\cite{Leggett74,vollhardtwolfle}
\begin{equation}
f_{\rm eff}=\frac{\mu_0\gamma^2}{2}{\bf S} \cdot
\stackrel{\leftrightarrow}{\chi}^{-1}\cdot{\bf S} 
- \gamma{\bf S}\cdot{\bf B}+f_{\rm d}+f_{\rm g} ,
\label{e.feff}
\end{equation}
where $\gamma$ is the gyromagnetic ratio and
$\stackrel{\leftrightarrow}{\chi}$ the susceptibility tensor. This
leads to the equations of motion 
\begin{subequations}\label{e.leggett}
\begin{eqnarray}
\dot{\bf S}&=&\gamma{\bf S}\times{\bf B}-\hat{\bf
d}\times\frac{\delta f}{\delta\hat{\bf d}}  ,   \\ 
\dot{\hat{\bf d}}&=&\hat{\bf d}\times\gamma\left({\bf
B}-\frac{\mu_0\gamma}{\chi}{\bf S}\right)  , 
\end{eqnarray}\end{subequations}
where $f=f_{\rm d}+f_{\rm g}$ and $\chi$ is the
susceptibility in the normal state. The motion of
$\hat{\bf l}$ is strongly limited by viscosity and therefore we assume
that $\hat{\bf l}$ is independent of time.\cite{LT2} Equations
(\ref{e.leggett}) describe dissipationless dynamics. The inclusion of
dissipative terms is postponed to Sec.~\ref{s.dissipation}. The field
${\bf B}$ is the sum of the static field ${\bf B}_0$ and a small 
radio-frequency field ${\bf B}'$ that oscillates at angular frequency
$\omega$. Throughout this paper we limit to study the  linear response
of ${\bf S}$ and $\hat{\bf d}$ to ${\bf B}'$. We parametrize the
deviation of $\hat{\bf d}$ with two parameters $d_\theta$ and $d_z$, 
\begin{eqnarray}
 \hat{\bf d} = \hat{\bf d}_0+(\hat{\bf z}\times \hat{\bf
d}_0)d_\theta+\hat{\bf z}d_z.
\label{e.drf}\end{eqnarray}
For ${\bf S}$  we parametrize the deviation ${\bf S}'$ from the
equilibrium ${\bf S}_0=\chi{\bf B}_0/\mu_0\gamma$ by $S'_z=S_z-S_0$
and circular components $S^{\pm}=S_x\pm\imagu S_y$. Similar
definitions are used for other vectors as well. We linearize Eqs.\
(\ref{e.leggett})  and assume the time dependence ${\bf S}'(t)={\bf
S}'\exp(-\imagu\omega t)$, etc. Using the equilibrium condition
(\ref{e.dtheta}) we get
\begin{subequations}\label{e.setcons}
\begin{eqnarray}
{\omega}S^{\pm} &=& \pm \omega_0S^{\pm}
\mp\lambda_{\rm d}e^{\pm\imagu\theta}
(\mathcal{D}+U_\perp)d_z
+\imagu \lambda_{\rm d}e^{\pm 2\imagu\theta}l_z l_\mp
d_\theta \mp \chi B_0B_\pm , \\
{\omega}d_z &=&
\frac{\mu_0\gamma^2}{2\chi}
\left(S^{-}e^{\imagu\theta}-S^{+}e^{-\imagu\theta}\right) +
\frac{\gamma}{2}
\left(B_{+}e^{-\imagu\theta}-B_{-}e^{\imagu\theta}\right) , \\
\omega S'_z &=&
-\imagu\lambda_{\rm d}(\mathcal{D}+U_\parallel) d_\theta
+\imagu\lambda_{\rm d}(\hat{\bf d}_0\times\hat{\bf l})_z l_z d_z
, \\
\omega d_\theta &=&
\imagu \frac{\mu_0\gamma^2}{\chi}S'_z -\imagu\gamma B'_z.
\end{eqnarray}
\end{subequations}
Here $\omega_0=\gamma B_0$ is the Larmor frequency.
The potentials $U_\parallel$ and $U_\perp$ are defined by
\begin{eqnarray}
U_\parallel &=& 1-l_z^2-2(\hat{\bf l}\times\hat{\bf d}_0)_z^2 
 \\
U_\perp &=&1-2 l_z^2-(\hat{\bf l}\times\hat{\bf d}_0)_z^2
-\frac{K_6}{\lambda_{\rm d}}
(\mbox{\boldmath$\nabla$}\theta)^2-\frac{K_5-K_6}{\lambda_{\rm
d}}(\hat{\bf l}\cdot\mbox{\boldmath$\nabla$}\theta)^2. 
\label{e.pot}
\end{eqnarray} 

In order to simplify Eqs.\ (\ref{e.setcons}), let us consider the
special case $l_z\equiv 0$. In this case the equations
separate into independent blocks for longitudinal and
transverse oscillations of the magnetization. The resonance
frequencies  are determined by two independent Schr\"odinger-type
equations for $d_\theta$ and $d_z$,
\begin{eqnarray}
(\mathcal{D}+U_\parallel) d_\theta &=&
\alpha_\parallel d_\theta , \label{e.schrod1}\\ 
(\mathcal{D}+U_\perp) d_z &=&
\alpha_\perp d_z .  
\label{e.schrod2}
\end{eqnarray}  
The eigenvalues $\alpha_{\parallel,k}$ and $\alpha_{\perp,k}$ of these
equations are related to the resonance frequencies as 
\begin{eqnarray}
\omega_{\parallel,k}^2 &=&
\Omega^2\alpha_{\parallel,k} , \label{e.freqll}\\
\omega_{\perp,k}^2 &=&
\omega_0^2 + \Omega^2\alpha_{\perp,k}.
\label{e.freqperpgen}
\end{eqnarray}
Here $\Omega = (\mu_0\gamma^2\lambda_{\rm d}/\chi)^{1/2}$ is the
longitudinal resonance frequency of the A phase. The corresponding
eigenfunctions of Eqs.\ (\ref{e.schrod1}) and (\ref{e.schrod2}) are
denoted by $\psi_{\parallel,k}$  and $\psi_{\perp,k}$,
respectively. They are assumed to be  normalized: 
$\int d^3r\,\vert\psi_{\parallel,k}\vert^2=1$ and
$\int d^3r\,\vert\psi_{\perp,k}\vert^2=1$. Because we have temporarily
neglected dissipative processes, the power absorption
$P(\omega)$ consists of $\delta$-peaks, $P(\omega)=\sum_k
I_k\delta(\omega-\omega_k)$.

Instead of assuming $l_z\equiv 0$, an alternative approach to Eqs.\
(\ref{e.setcons}) is to study the high-field limit
$\omega_0\gg\Omega$. More precisely, one can calculate the resonance
frequencies as a power series of $\Omega^2$ and neglect terms of the
order of $\Omega^4$ and higher. In this approximation all the three
components $S'_z$, $S^+$, and $S^-$ decouple. The eigenvalue equations
and frequencies are the same as above equations
(\ref{e.schrod1})-(\ref{e.freqperpgen}) except that Eq.\
(\ref{e.freqperpgen}) is valid only to leading order in $\Omega$:  
\begin{eqnarray}
\omega_{\perp,k}
&=&\pm\left[ \omega_0 +
\frac{\Omega^2}{2\omega_0}\alpha_{\perp,k}
+O(\frac{\Omega^4}{\omega_0^3})\right].
\label{e.freqperp}
\end{eqnarray} 
For the rest of this section we assume the high-field
limit $\omega_0\gg\Omega$.

In the case of dipole locking, $\hat{\bf l}({\bf r}) =
\hat{\bf d}({\bf r})$, the lowest bulk eigenvalues are
$\alpha_{\rm \parallel,b} = \alpha_{\rm \perp,b} = 1$. In this case
only the bulk eigenstate gives rise to absorption 
$I_{{\rm b},\parallel}=Vi_\parallel$ and $I_{{\rm b},\perp}=Vi_\perp$,
where the two modes 
\begin{eqnarray}
i_{\parallel}&=& \frac{\pi}{4\mu_0} {B'_z}^2 
\chi  \Omega^2 , \\
i_{\perp}&=& \frac{\pi}{4\mu_0} (B_x^2+B_y^2) \chi \omega_0^2, 
\label{e.abspower}
\end{eqnarray} 
and $V=\int d^3r$ is the volume of the liquid. 

In the presence of dipole unlocking also other eigenstates contribute
to the absorption. Their intensities are given
by\cite{BruinsmaMaki,VSF84,vollhardtwolfle} 
\begin{eqnarray}
I_{\parallel,k} &=& i_{\parallel}\,\alpha_{\parallel,k}\,
Q_{\parallel,k} = i_{\parallel}\,\alpha_{\parallel,k}
\left\vert\int d^3r
\psi_{\parallel,k}({\bf r})\right\vert^2  \label{e.intensity1} , \\
I_{\perp,k} &=&i_{\perp} Q_{\perp,k} =
i_{\perp}\left\vert\int d^3r\psi_{\perp,k}({\bf r})\exp[{\rm
i}\theta({\bf r})]\right\vert^2. 
\label{e.intensity2}
\end{eqnarray} 
Here the $Q_k$'s satisfy the sum rules
\begin{eqnarray}
\sum_{k=0}^{\infty} Q_{\parallel,k} &=&\sum_{k=0}^{\infty}
Q_{\perp,k} = V  \label{e.sumrule0} , \\
\sum_{k=0}^{\infty} \alpha_{\parallel,k} Q_{\parallel,k} &=& \int d^3r\,
U_\parallel   \label{e.sumrule1a} , \\
\sum_{k=0}^{\infty} \alpha_{\perp,k} Q_{\perp,k} &=& \int d^3r
\left[(\hat{\bf l} \cdot \hat{\bf d}_0)^2 - l_z^2 \right] 
\label{e.sumrule1b} , \\
\sum_{k=0}^{\infty} \alpha_{\parallel,k}^2 Q_{\parallel,k} &=& \int
d^3r\, U_\parallel^2  \label{e.sumrule2a} , \\
\sum_{k=0}^{\infty} \alpha_{\perp,k}^2 Q_{\perp,k} &=& \int
d^3r\left\{
 \left[(\hat{\bf l} \cdot \hat{\bf d}_0)^2 - l_z^2\right]^2 +
(\hat{\bf l}\times\hat{\bf d}_0)_z^2 (\hat{\bf l} \cdot \hat{\bf
d}_0)^2 \right\} ,
\label{e.sumrule2b}
\end{eqnarray}
and so on. The sum rules can be derived using the orthogonality
properties of the eigenfunctions. [In Eq.\ (\ref{e.sumrule2b}) one also
needs Eq.\ (\ref{e.dtheta}).] The lowest-order rules
(\ref{e.sumrule0})-(\ref{e.sumrule1b}) apparently are equivalent to the
sum rules presented by Leggett.\cite{Leggett73}

\section{EQUILIBRIUM STRUCTURE OF A SPLAY SOLITON}\label{s.eqstruc}

\begin{figure}[!tb]
\begin{center} 
\includegraphics[width=8cm]{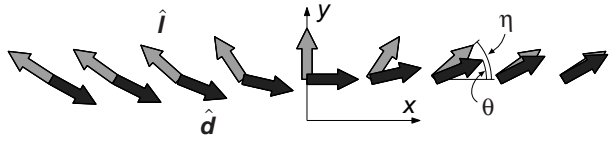}
\caption{The structure of a splay soliton, where the $\hat{\bf l}$
field (light arrows) has splay shape. The structure is homogeneous
in the $y$-$z$ plane of the soliton wall. Both $\hat{\bf d}$ and
$\hat{\bf l}$ are perpendicular to the static field ${\bf B}_0 = B_0
\hat{\bf z}$.}  
\label{splay} 
\end{center}
\end{figure}

The minimum of the dipole energy (\ref{e.fdip}) can
be achieved in two ways: either $\hat{\bf l}$ and $\hat{\bf d}$ are
parallel or they are antiparallel. This double degeneracy gives
rise to domain walls or solitons. On one side of the wall $\hat{\bf l}
= \hat{\bf d}$ and on the other side $\hat{\bf l}=-\hat{\bf d}$. We
will study cases where the static
field ${\bf B}_0$ is either perpendicular or parallel
to the plane of the soliton wall. The soliton structures in these
cases are known to have composite twist and splay
structures, respectively.\cite{MakiKumar} The structure and the
fundamental NMR frequency of a twist soliton have  been solved
analytically.\cite{MakiKumar77,MakiKumar,vollhardtwolfle} Here we
concentrate on the case of a splay soliton, which has previously
been studied only by variational
methods.\cite{MakiKumar,VollhardtMaki,BruinsmaMaki,exp} Its structure
is illustrated in Fig.~\ref{splay}. 

\begin{figure}[!tb]
\begin{center} 
\includegraphics[width=8cm]{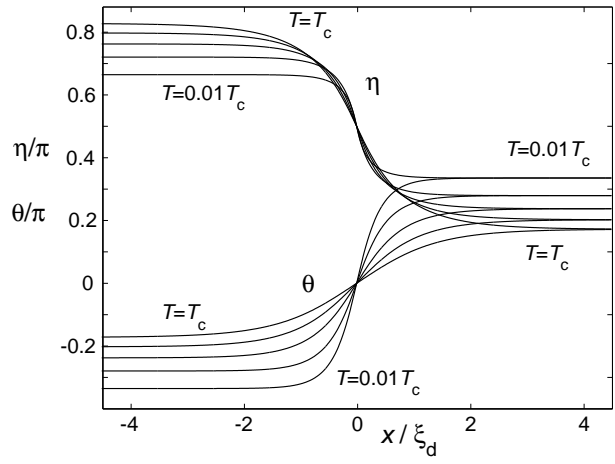}
\caption{The functions $\eta(x)$ and $\theta(x)$ for a splay soliton 
at temperatures $T/T_{\rm c} = 0.01, 0.25, 0.50, 0.75, 1.00$. The
parameter values correspond to weak coupling at 3.4 MPa ($F_1^s=14.45$
and $F_1^a=-1$).}   
\label{angles} 
\end{center}
\end{figure}

In the cases of a twist or a splay soliton, both $\hat{\bf l}$ and
$\hat{\bf d}$  are in the plane perpendicular to ${\bf B}_0$.
We parametrize 
\begin{eqnarray} 
\hat{\bf l}=\hat{\bf x}\cos\eta+\hat{\bf y}\sin\eta  
\label{e.ldvect}
\end{eqnarray} 
and $\hat{\bf d}$ as above, Eq.\ (\ref{e.d0}). For a splay soliton the
angles $\theta$ and $\eta$ are taken as functions of $x$ only. We use
the boundary conditions $\eta(+\infty)=\theta(+\infty)$ and
$\eta(-\infty)=\theta(-\infty)+\pi$. Substituting the vector fields
(\ref{e.d0}) and (\ref{e.ldvect}) into the total energy
(\ref{e.fstatic}) gives that the energy per unit area is    
\begin{eqnarray} 
f_{\rm s}&=&\frac{1}{2}\int dx [(K_{\rm s}\sin^2\eta
+ K_{\rm b}\cos^2\eta)\left(\frac{d\eta}{dx}\right)^2 \nonumber \\
&& + (K_6\sin^2\eta +K_5\cos^2\eta)\left(\frac{d\theta}{dx}\right)^2 -
\lambda_{\rm d}\cos^2(\eta-\theta) ] \ .  
\label{e.fcsplay0}
\end{eqnarray}
The energy functional (\ref{e.fcsplay0}) was discretized using equal
intervals ($x_i=-L/2+Li/N$, $i=0, \ldots, N$), and the values
of $\eta$ and $\theta$ at these discrete points were taken to be
the minimizing variables. The boundary conditions were taken into
account by linear initial approximations $\eta = (\pi/2)-
(2\pi/3L)x$ and $\theta = (\pi/3L)x$. The
minimization can be done by a simple relaxation method. The
resulting functions $\eta(x)$ and $\theta(x)$ at different
temperatures are shown in Fig.~\ref{angles}. 

\section{PRINCIPAL NMR FREQUENCY}\label{s.princnmr}

We now apply the theory of Sec.\ \ref{s.hyrdo} to calculate NMR
properties of solitons. For solitons $l_z \equiv 0$, which
decouples the longitudinal and transverse modes at arbitrary $\omega_0$
in Eqs.\ (\ref{e.setcons}).  Thus the eigenvalues $\alpha_{\perp,k}$ are
related to resonance frequencies as given in Eq.\   
(\ref{e.freqperpgen}) at any field. 

\begin{figure}[!tb] 
\begin{center} 
\includegraphics[width=8cm]{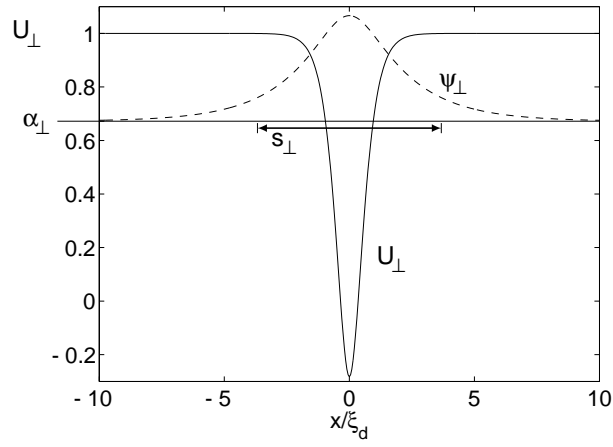}
\caption{The potential $U_\perp$ (solid line), the lowest eigenvalue
$\alpha_\perp$, and the eigenfunction $\psi_\perp$ (dashed line,
vertical scale arbitrary) for transverse resonance of a splay soliton
at $T\approx T_{\rm c}$. The absorption thickness $s_\perp$ gives the
effective width of the wave function in NMR.  } 
\label{wavefunc} 
\end{center}
\end{figure}

\begin{figure}[!tb] 
\begin{center} 
\includegraphics[width=8cm]{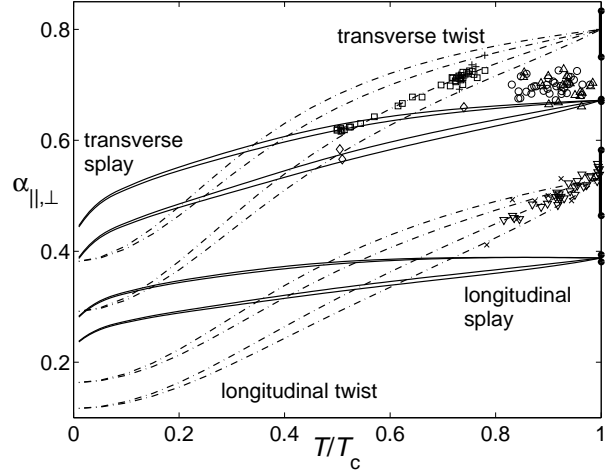}
\caption{The reduced resonance frequencies $\alpha_\parallel$
(\ref{e.freqll}) and $\alpha_\perp$ (\ref{e.freqperpgen}) as
functions of temperature. Splay-soliton transverse resonance: four
upper solid lines and experimental points $\circ$ (Ref.~\onlinecite{GouldLee}),
$\triangle$ (Ref.~\onlinecite{GouldBartolacBozler}), and
$\diamondsuit$ (Ref.~\onlinecite{exp}). Splay-soliton longitudinal
resonance: four lower solid lines. Twist-soliton transverse resonance:
four upper dash-dotted lines and experimental points $+$
(Ref.~\onlinecite{Hakonen}) and $\square$ (Ref.\onlinecite{exp}). 
Twist-soliton longitudinal resonance: four lower dash-dotted lines and
experimental points $\times$ (Ref.~\onlinecite{Avenel}) and
$\triangledown$ (Ref.~\onlinecite{GouldLee}). In each case there are four
theoretical lines, which in order of decreasing $\alpha$ correspond to
$F_1^a = 0$ \& weak coupling, $F_1^a = 0$ \& gap enhancement,
$F_1^a = -1$ \& weak coupling, and $F_1^a = -1$ \& gap enhancement.
Good agreement between theory and experiment is achieved with $F_1^a =
-1$ and weak coupling except for the high-temperature transverse splay
data. The effect of strong-coupling corrections is shown at
$T/T_{\rm c}=1$ where the lower ends of the bars correspond to
$\gamma=2$ and the upper ends to $\gamma=4$. The experimental data is
plotted with higher resolution in Fig.\ \protect\ref{r2spindiff}.    } 
\label{R2} 
\end{center}
\end{figure}

The main signature of solitons in the
NMR spectrum comes from the lowest eigenvalue of  Schr\"odinger-like
equations (\ref{e.schrod1}) and (\ref{e.schrod2}). This lowest
frequency can be calculated, for example, using a variational
formulation: 
\begin{eqnarray}
\alpha_{\parallel,\perp}&=& \min_\psi \frac{\int d^3r\left\lbrack
K_6\vert\mbox{\boldmath$\nabla$}\psi\vert^2 +(K_5-K_6)\vert\hat{\bf
l}\cdot\mbox{\boldmath$\nabla$}\psi\vert^2 + \lambda_{\rm d}
 U_{\parallel,\perp}
\vert\psi\vert^2 \right\rbrack }{ \lambda_{\rm d}\int
d^3r\vert\psi\vert^2 } .
\label{e.varform}
\end{eqnarray}
This was discretized and the values of $\psi(x_i)$ were taken to be
the minimizing variables. The form for the initial approximation used
was $\psi(x)=\cosh^n(qx)$ and  
the boundary conditions ${d\psi}/{dx}=0$ were assumed at the end
points. The length $L$ was increased until its effect disappeared. 
One solution is shown in Fig.\ \ref{wavefunc}.
In Fig.\ \ref{R2} we plot the temperature dependence of the lowest
resonance frequencies as well as some experimental data. For
completeness we  also include the analytical results for the twist
soliton:\cite{MakiKumar,vollhardtwolfle} 
\begin{eqnarray}
\alpha_\parallel^{\rm twist}&=&\frac{1}{2K_{\rm t}} \lbrack
\sqrt{(9K_{\rm t}+K_6)(K_{\rm t}+K_6)}-3K_{\rm t}-K_6\rbrack \\
\alpha_\perp^{\rm twist}&=&\frac{K_6}{K_6+K_{\rm t}} .
\label{R2twist}
\end{eqnarray}

The reduced frequency shifts $\alpha_\parallel$ (\ref{e.freqll}) and
$\alpha_\perp$ (\ref{e.freqperpgen}) depend only on the ratios of
the hydrodynamic coefficients $K_{\rm b}$,
$K_{\rm s}$, $K_{\rm t}$, $K_{5}$, and $K_{6}$. (Note that the absolute
magnitudes of $K_i$'s and $\lambda_{\rm d}$ define length and energy
scales that do not affect $\alpha_\parallel$ or $\alpha_\perp$.)  In
the weak-coupling approximation the ratios of $K_i$'s are functions of
an infinite set of Fermi-liquid parameters $F_j^s$ and $F_j^a$, with
$j= 1, 3, 5$, etc.\cite{Dorfle} Here we neglect all the coefficients
with $j>1$ since they are unknown. For $F_1^s$ we use the value
by Greywall\cite{Greywall} at the melting pressure, $F_1^s=14.5$.
However, the dependence on $F_1^s$ is weak. For example, the variation 
$F_1^s=14.5\pm 1$  shows up only in the fourth decimal of $\alpha$
at $T=0.8T_{\rm c}$. In terms of pressure, the maximum
difference in $\alpha$ between 2.6 and 3.4 MPa is 1\% at
temperatures higher than $0.5T_{\rm c}$. The value
$F_1^a=-1$ was taken from Ref.~\onlinecite{Greywall83} and it is also
consistent with Ref.~\onlinecite{Candela}. 
In order to see the effect of $F_1^a$ we
also used $F_1^a=0$. It can be seen in Fig.\ \ref{R2} that this shifts
the resonance frequency up at temperatures below $T_{\rm c}$. 

There are no quantitative calculations of strong-coupling effects in the
A phase at a general temperature. In order to get some idea how strong
coupling could affect the soliton frequencies, we use a ``trivial
strong-coupling'' model developed by Serene and
Rainer\cite{SereneRainer} for the B phase. In this model the
weak-coupling energy gap is multiplied by a factor that depends on the
temperature and on the jump of the specific heat $\Delta C_{\rm
B}/C_{\rm n}$ at $T=T_{\rm c}$. This dependence is tabulated in Ref.\
\onlinecite{SereneRainer}. We adapt this model to the A phase by
calculating the multiplying factor using the same table but substituting
$\frac{6}{5}\Delta C_{\rm A}/C_{\rm n}$ in place of
$\Delta C_{\rm B}/C_{\rm n}$, and using extrapolation when needed. We
take $\Delta C_{\rm A}/C_{\rm n}$ from measurements by 
Greywall.\cite{Greywall} It can be seen in
Fig.\ \ref{R2} that the gap enhancement, which affects only
intermediate temperatures, has a smaller effect than the change of
$F_1^a$. 

In the limiting case $T\rightarrow T_{\rm c}$ the reduced frequencies
$\alpha_{\parallel}$ and
$\alpha_{\perp}$ are independent of any parameters appearing
in the weak-coupling model, including also the gap enhancement. This is
a consequence of the Ginzburg-Landau expansion that gives to the
parameters $K_i$ the ratios $K_{\rm b} : K_{\rm s} : K_{\rm t} : K_5 :
K_6 = \gamma : 1 : 1 : 2 : (\gamma+1)$. The only free parameter here is
$\gamma$, which in the weak coupling (with or without gap
enhancement) has the value $\gamma = 3$. This value is
changed only when nontrivial strong-coupling corrections are included.
The ``weak-coupling-plus'' model by Serene and
Rainer\cite{SereneRainer2} gives an estimate $\gamma \approx 3.12$. 
Figure \ref{R2} shows the reduced frequencies corresponding to
$\gamma=2$ and $\gamma=4$. We see that this variation changes the
splay soliton $\alpha_\perp$ less than 3\% but for the twist soliton
the effect is ten times larger.

There is rather good agreement between the experiment and the theory
corresponding to weak coupling and $F_1^a=-1$. Equally good agreement
is achieved with gap enhancement and $F_1^a\approx -0.7$. The
longitudinal twist data is a strong indication that the deviation from
the weak-coupling value $\gamma=3$ is small, as predicted by Serene and
Rainer.\cite{SereneRainer2} The only major difference between theory
and experiment exists in the transverse splay-soliton frequency at high
temperatures. It seems very difficult to improve the agreement in this
case by any change of the parameters in the theory above since
$\alpha_\perp^{\rm splay}$ at
$T_{\rm c}$ is effectively fixed. On the experimental side, one
possibility is that the field is not precisely in the plane of the
soliton in the measurements. This would add a small twist component to
the splay soliton and thus shift up the frequency.\cite{MakiKumar}
Another possibility is that the inclusion of relaxation mechanisms
could shift the calculated resonance frequency, as will be discussed in
Sec.\ \ref{s.dissipation}.

At $T_{\rm c}$  we find the eigenvalues
$\alpha_\parallel=0.388$ and $\alpha_\perp=0.672$ for $\gamma=3$.
These differ slightly from the variational results by Maki and
Kumar\cite{MakiKumar} that are 0.403 and 0.677, respectively.

\section{ABSORPTION AND HIGHER MODES}\label{s.intes}

\begin{figure}[!tb]
\begin{center} 
\includegraphics[width=8cm]{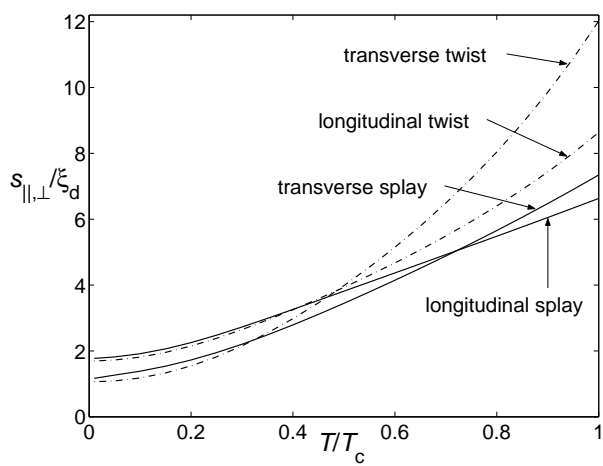}
\caption{Absorption (\protect\ref{e.abslendef}) of the principal soliton
peak in four cases. The parameter values correspond to weak coupling
at 3.4 MPa ($F_1^s=14.45$ and $F_1^a=-1$). 
The numerical values at $T_{\rm c}$ are $s_\parallel^{\rm splay} =
6.63\xi_{\rm d}$, $s_\perp^{\rm splay} = 7.35\xi_{\rm d}$,
$s_\parallel^{\rm twist}=8.64\xi_{\rm d}$, and $s_\perp^{\rm
twist}=12.02\xi_{\rm d}$.  }   
\label{abslen} 
\end{center}
\end{figure}

Here we calculate the intensity of the principal soliton peak and
analyze the absorption at other frequencies. For planar objects
intensities (\ref{e.intensity1}) and (\ref{e.intensity2}) are most
conveniently expressed in the form of an absorption thickness $s_{k}$,
which equals $Q_{k}$ divided by the area of the planar object,
\begin{eqnarray}
s_{k} = \frac{Q_{k}}{A},
\label{e.abslendef}
\end{eqnarray} 
for each mode $k$.  The absorption thicknesses
in the lowest eigenstate are plotted in Fig.~\ref{abslen}.

The absorption for a twist soliton can be calculated
analytically. Substituting
the eigenfunction $\psi=c_1\mathop{\rm sech}^\mu az$ into 
Eqs.\ (\ref{e.intensity1}) and (\ref{e.intensity2}) 
and using $\theta=c_2-\mu_\perp\mathop{\rm
sgn}(z)\arccos(\mathop{\rm sech} az)$ (Refs.\
\onlinecite{MakiKumar} and \onlinecite{vollhardtwolfle}) we find
\begin{eqnarray}
s_\parallel^{\rm twist} &=& 
\frac{ \sqrt{\pi}\Gamma^2(\frac{\mu_\parallel}{2})
\Gamma(\mu_\parallel+\frac{1}{2}) }{a \Gamma(\mu_\parallel) 
\Gamma^2(\frac{\mu_\parallel+1}{2})} \label{st1}\\  
s_\perp^{\rm twist} &=&
\frac{\sqrt{\pi}4^{1-\mu_\perp}\Gamma(\mu_\perp)}
{a\Gamma(\mu_\perp+\frac{1}{2})}. \label{st2}
\end{eqnarray}
Here $\Gamma$ is the Gamma function and 
\begin{eqnarray}
a&=&\sqrt{\frac{\lambda_{\rm d}(K_6+K_{\rm t})}{K_6 K_{\rm t}}}\\
\mu_\parallel&=&\frac{1}{2}\left(\sqrt{\frac{9K_{\rm t}+K_6}{K_{\rm
t}+K_6}} -1 \right)  \\
\mu_\perp&=&\frac{K_{\rm t}}{K_6+K_{\rm t}} .
\label{e.mudef}\end{eqnarray}

In order to get all the resonance modes, we discretize the space as
discussed above ($x_i=-L/2+Li/N$, $i=0, \ldots, N$, for twist soliton
replace $x$ by $z$). This means that Eqs. (\ref{e.schrod1}) and
(\ref{e.schrod2}) turn into a matrix eigenvalue problem. This can be
solved by standard library routines for matrices of reasonable size.
This gives all the frequencies, including the fundamental one
(\ref{e.varform}). However, the higher modes extend over the whole
interval, and it is important to fix boundary conditions for them. We
require zero derivative of $\psi$ at $x=\pm L/2$. This can be justified
by considering a lattice of solitons. The general eigenfunctions $\psi$
of Eqs.\ (\ref{e.schrod1}) and (\ref{e.schrod2}) are of the Bloch form,
but only strictly periodic functions lead to nonzero absorption in Eqs.\
(\ref{e.intensity1}) and (\ref{e.intensity2}).  The unit cell of a
soliton lattice $x=-L/2, \ldots, 3L/2$ consists of two solitons located
around $x=0$ and $x=L$.  The vectors $\hat{\bf l}$ and
$\hat{\bf d}$ have the symmetries $\hat{\bf l}(L/2+x) = \hat{\bf
l}(L/2-x)$ and $\hat{\bf d}(L/2+x) = \hat{\bf d}(L/2-x)$. This implies
that the eigenfunctions can be classified as symmetric or
antisymmetric (with respect to $x=L/2$). The symmetric solutions have
$d\psi/dx(\pm L/2)=0$, and the antisymmetric solutions can be
neglected since they do not contribute to the absorption. 

\begin{figure}[!tb] 
\begin{center} 
\includegraphics[width=8cm]{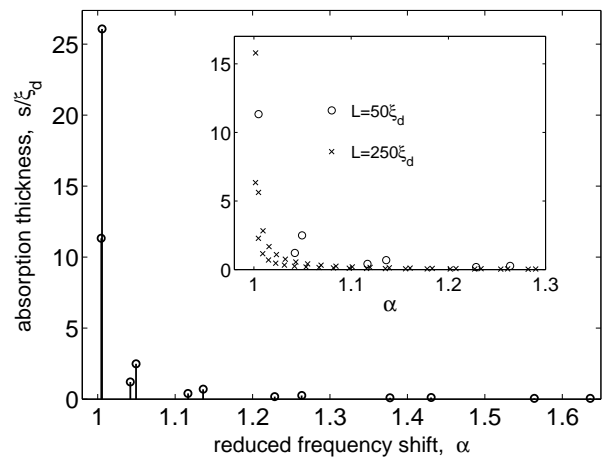}
\caption{The higher modes in the case of transverse resonance of a
splay soliton.  The length used in the calculation is
$L=50\xi_{\rm d}$ in the main frame whereas the inset also contains
the points for $L=250\xi_{\rm d}$. In both cases the lattice spacing
of the discretized lattice is $0.1\xi_{\rm d}$. Other parameters are
weak coupling at 3.4 MPa ($F_1^s=14.45$, $F_1^a=-1$) and temperature
$T=0.95T_{\rm c}$.}
\label{highmode} 
\end{center}
\end{figure}

The results for the frequencies and intensities of the higher modes are
shown in Fig.~\ref{highmode}.
As expected, the higher modes depend on the size $L$ of the system
used in the calculation. The modes seem to appear as pairs above the
bulk peak. The bulk peak also seems to consist of two peaks that do
not appear exactly at $\alpha = 1$ due to finite $L$. In fact, the
finite $L$ corresponds to a periodic lattice of solitons, which should
give rise to peaks that are analogous to the Bragg peaks in the
x-ray scattering from periodic solids. Such peaks are indeed seen
experimentally in the case of a vortex sheet, where a nearly periodic
arrangement is automatically generated.\cite{Parts} 
The observation of the peaks depends essentially on dissipative
effects, which broadens the peaks. This will be discussed in the
following section.

\section{THE EFFECT OF DISSIPATION}\label{s.dissipation}

There are two sources of dissipation that are important for solitons.
The nonequilibrium between normal and superfluid magnetizations is
the main relaxation mechanism in homogeneous superfluid
$^3$He.\cite{CE,Combescot76,LT} In inhomogeneous situations the
nonequilibrium between normal magnetizations at different locations
causes relaxation via spin diffusion. We study these two mechanisms
below, the latter in the case where the inhomogeneity arises from the
soliton. We neglect extrinsic effects such as the
inhomogeneity caused by nonuniform magnetic field. We also do not
consider dissipation caused by orbital motion. According to the
estimation in Ref.\ \onlinecite{LT2}, this is important only at
temperatures very near $T_{\rm c}$, region which is vanishingly small
on the scale of Fig.\ \ref{R2}.

Our treatment of the dissipation is purely phenomenological. For
normal-superfluid relaxation we use the model developed by Leggett and
Takagi, which is extensively described in
Ref.~\onlinecite{LT}. For the relaxation of spatially nonequilibrium
magnetization we use simple diffusion equation.  This approach can be
correct only when the mean free path of the quasiparticles is smaller
than the thickness of the soliton, and therefore it necessarily fails at
low temperatures, where the mean free path diverges. Our treatment also
neglects the tensor character of the spin-diffusion coefficient, as
well as the anisotropic energy gap with varying anisotropy axis
$\hat{\bf l}({\bf r})$, which leads to  Andreev reflection of the
quasiparticles.

In order to treat the relaxation, the spin
polarization ${\bf S}$ is divided into a superfluid (pair) part ${\bf
S}_p$ and a normal (quasiparticle) part ${\bf S}_q= {\bf S}-{\bf
S}_p$. Both parts have separate equations of motion   
\begin{subequations}\label{e.ltequ2}\begin{eqnarray}
\dot{\bf S}_q &=& \gamma{\bf S}_q \times \left( {\bf B} -
\mu_0\gamma\frac{F_0^a}{\chi_{0}} {\bf S}_p \right) +
\frac{1}{\tau}\left[ (1-\lambda) {\bf S}_p - \lambda {\bf S}_q \right]
+ \kappa \nabla^2 {\bf S}_q\label{e.ltequ2a}
 \\ 
\dot{\bf S}_p &=& \gamma{\bf S}_p \times \left( {\bf B} -
\mu_0\gamma \frac{F_0^a}{\chi_{0}} {\bf S}_q \right) -
\frac{1}{\tau}\left[ (1-\lambda) {\bf S}_p - \lambda {\bf S}_q \right]
-\hat{\bf d}
\times \frac{\delta f}{\delta\hat{\bf d}}  \\ 
\dot{\hat{\bf d}} &=& \gamma\hat{\bf d}\times \left[ {\bf B} -
\mu_0\gamma\frac{F_0^a}{\chi_{0}}{\bf S}_q
-\mu_0\gamma
\left(\frac{F_0^a}{\chi_{0}}+\frac{1}{\lambda\chi_0}\right)
{\bf S}_p
\right] .  
\end{eqnarray}\end{subequations}
Except the spin-diffusion term $\kappa \nabla^2 {\bf
S}_q$, these equations are the same as in Ref.\ \onlinecite{LT}. Here
$\chi_0$ is the susceptibility in the absence of Fermi-liquid effects.
As above, $\chi$ is the susceptibility in the normal state, so that
$\chi/\chi_0 = 1/(1+F_0^a)$. The function $\lambda(T/T_{\rm c})$ is
defined as the equilibrium fraction of the superfluid magnetization,
and is given by Eq.\ (4.23) in Ref.~\onlinecite{LT}. The Leggett-Takagi
relaxation time $\tau$ describes local relaxation of ${\bf
S}_p$ and ${\bf S}_q$ to their equilibrium values. 
The spin diffusion appears via term $\kappa \nabla^2 {\bf S}_q$ in the
equation for $\dot{\bf S}_q$ (\ref{e.ltequ2a}). In the normal state
the spin-diffusion constant $\kappa$ is related to the spin-diffusion
time $\tau_{\rm D}$ by\cite{Leggett70}  
\begin{equation}
\kappa = \frac{1}{3}v_{\rm F}^2 (1+F_0^a)\tau_{\rm D},
\label{e.diffcoeff}
\end{equation}
where $v_{\rm F}$ is the Fermi velocity. We use this parametrization
also in the superfluid state. Note that $\kappa$ is defined as the
spin-diffusion constant for ${\bf S}_q$, so that the effective diffusion
constant for the total magnetization ${\bf S}$ is $(1-\lambda)\kappa$.

We continue by writing the equations for variables $\hat{\bf d}$, ${\bf
S}$, and the deviation from local equilibrium
\begin{eqnarray}
\boldsymbol{\eta} = {\bf S}_p - \lambda {\bf S}.
\end{eqnarray}   
Similarly, as in Sec.\ \ref{s.hyrdo}, we linearize the equations and
assume harmonic time dependence. We use the fact that 
$l_z=0$ in solitons. As a consequence, the transverse and longitudinal
modes separate at any field $B_0$. Differing from Sec.\
\ref{s.hyrdo} we introduce dimensionless quantities by defining
$S^\pm = (\Omega^2/\lambda_{\rm d} \omega_0) (S_x' \pm \imagu S_y')$ and
$\eta^\pm = (\Omega^2/\lambda_{\rm d} \omega_0)(\eta_x' \pm \imagu
\eta_y')$. For transverse oscillations we get the equations 
\begin{subequations}\label{e.eigvgen}\begin{eqnarray}
\frac{\omega}{\omega_0}S^{\pm} &=& 
\left[\pm 1+\imagu (1-\lambda)\frac{\kappa}{\omega_0}\nabla^2\right]S^{\pm}
- \imagu\frac{\kappa}{\omega_0}\nabla^2\eta^{\pm}
\mp\frac{\Omega^2}{\omega_0^2}e^{\pm \imagu\theta}
(\mathcal{D}+U_\perp)d_z
\mp \frac{B_\pm}{B_0} \label{e.eigvgena} , \\
\frac{\omega}{\omega_0}\eta^{\pm} &=& 
\left(\pm \frac{1}{1+F_0^a}-\frac{\imagu}{\omega_0\tau} +
\imagu\lambda\frac{\kappa}{\omega_0}\nabla^2\right)\eta^{\pm} 
-\imagu\lambda(1-\lambda)\frac{\kappa}{\omega_0}\nabla^2 S^{\pm}
\nonumber  \\
&&\mp(1-\lambda)\frac{\Omega^2}{\omega_0^2}e^{\pm\imagu\theta}
(\mathcal{D}+U_\perp)d_z
\label{e.eigvgenb} , \\
\frac{\omega}{\omega_0}d_z &=&
\frac{1}{2}\left(S^{-}e^{\imagu\theta}-S^{+}e^{-\imagu\theta}\right)
+\frac{\chi}{2\lambda\chi_0}\left(\eta^{-}e^{\imagu\theta} -
\eta^{+}e^{-\imagu\theta}\right) 
\nonumber \\ && + \frac{1}{2B_0}
\left(B_{+}e^{-\imagu\theta}-B_{-}e^{\imagu\theta}\right).
\end{eqnarray}\end{subequations}
The full solution of this problem can be written as a five-component
vector $\psi = (S^{-}\ \eta^{-}\ d_z\ \eta^{+}\ S^{+})^T =
\sum_i c_i \psi_i $, where the $\psi_i$'s satisfy the homogeneous
equation where $B_\pm=0$. The eigenvalues $\omega_i$ of
the homogeneous equation are now complex valued, and
the eigenvectors $\psi_i$ are not orthogonal to each other. In order
to solve for the absorption spectrum we find that the following
adjoint eigenvalue problem:\cite{MorseF} 
\begin{subequations}\label{e.eigvgen2}
\begin{eqnarray}
\frac{\omega}{\omega_0}\underline{S}^{\pm} &=& 
\left[\pm 1+\imagu (1-\lambda)\frac{\kappa}{\omega_0}\nabla^2\right]
\underline{S}^{\pm}
-\imagu\lambda(1-\lambda)\frac{\kappa}{\omega_0}\nabla^2 
\underline{\eta}^{\pm}
\mp\frac{1}{2}e^{\mp\imagu\theta}\underline{d}_z
 \\
\frac{\omega}{\omega_0}\underline{\eta}^{\pm} &=& 
\left(\pm \frac{1}{1+F_0^a}-\frac{\imagu}{\omega_0\tau} +
\imagu\lambda\frac{\kappa}{\omega_0}\nabla^2\right)\underline{\eta}^{\pm} 
- \imagu\frac{\kappa}{\omega_0}\nabla^2 \underline{S}^{\pm}
\mp\frac{\chi}{2\lambda\chi_0} e^{\mp\imagu\theta}\underline{d}_z
 \\
\frac{\omega}{\omega_0}\underline{d}_z &=&
\frac{\Omega^2}{\omega_0^2}(\mathcal{D}+U_\perp)\left[
e^{-\imagu\theta}\underline{S}^{-} - e^{\imagu\theta}\underline{S}^{+}
+(1-\lambda)(e^{-\imagu\theta}\underline{\eta}^{-} -
e^{\imagu\theta}\underline{\eta}^{+}) \right]
\end{eqnarray}\end{subequations}
has the same eigenvalues $\omega_i$ as the homogeneous version of
problem (\ref{e.eigvgen}) and that the eigenvectors
$\underline{\psi}_i=(\underline{S}_i^{-}\ \underline{\eta}_i^{-}\
\underline{d}_{z,i}\ \underline{\eta}_i^{+}\ \underline{S}_i^{+})^T$
are orthogonal to $\psi_j$'s: 
\begin{eqnarray} 
\int \underline{\psi}_i^T \psi_j d^3r = \delta_{ij}.
\label{e.ortog}
\end{eqnarray} 
When deriving this we have assumed zero derivate for the wave
functions far from the soliton by considering a lattice of solitons as
was discussed in Sec.\ \ref{s.intes}. The power absorption can now
be written as
\begin{eqnarray}
P(\omega) &=& \frac{1}{2}\gamma\omega\int {\rm Im}\left( {\bf
B}'\cdot {\bf S}' \right) d^3r \nonumber \\ 
&=& \frac{\chi B_0 \omega}{4\mu_0} \sum_j {\rm Im}
\left[ c_j \int d^3r \left( B_{-}^{*} S_j^{-}+B_{+}^{*} S_j^{+}
\right) \right] ,
\label{e.powerabs}
\end{eqnarray}
where the coefficients $c_j$ are given by
\begin{eqnarray}
c_j = \frac{\gamma}{\omega-\omega_j} \int d^3r \left[
 B_{-}\underline{S}_j^{-} -B_{+}\underline{S}_j^{+}
+ \frac{1}{2}(e^{-\imagu\theta}B_{+} -
e^{\imagu\theta}B_{-})\underline{d}_{z,j} \right] . 
\end{eqnarray}
We observe that the eigenfunctions are not symmetric with respect
to the center of the soliton. This apparently is caused by the
spin-diffusion term.

In the hydrodynamic limit where $\omega\tau \ll 1$ one may solve
$\eta^\pm$ from Eq. (\ref{e.eigvgen}) and to linear order in $\tau$
\begin{eqnarray}
\eta^\pm = -\tau\lambda(1-\lambda)\kappa\nabla^2 S^\pm
\pm \imagu\tau(1-\lambda)\frac{\Omega^2}{\omega_0}e^{\pm\imagu\theta}
(\mathcal{D}+U) d_z .
\end{eqnarray}
If we additionally ignore the term $\nabla^2\eta^\pm$, the equations
for the three-component vector $(S^{-}\ d_z\ S^{+})^T$ read 
\begin{subequations}\label{e.eigvhydr}
\begin{eqnarray}
\frac{\omega}{\omega_0}S^{\pm} &=& 
\left[\pm 1+\imagu (1-\lambda)\frac{\kappa}{\omega_0}\nabla^2\right]S^{\pm}
\mp\frac{\Omega^2}{\omega_0^2}e^{\pm
\imagu\theta}(\mathcal{D}+U_\perp)d_z
\mp \frac{B_\pm}{B_0} , \\
\frac{\omega}{\omega_0}d_z &=&
\frac{1}{2}e^{\imagu\theta}
\left[1-\frac{\tau\chi(1-\lambda)\kappa}{\chi_0}
\nabla^2\right] S^{-} 
-\frac{1}{2}e^{-\imagu\theta}
\left[1-\frac{\tau\chi(1-\lambda)\kappa}{\chi_0}
\nabla^2\right] S^{+} \nonumber \\
&-&\imagu\frac{\tau\chi\Omega^2(1-\lambda)}{\chi_0\omega_0\lambda} 
(\mathcal{D}+U_\perp)d_z
+ \frac{1}{2B_0}
\left(B_{+}e^{-\imagu\theta}-B_{-}e^{\imagu\theta}\right) .
\end{eqnarray}\end{subequations}
The homogeneous adjoint problem for this is given by 
\begin{subequations}\label{e.eigvhydr2}
\begin{eqnarray}
\frac{\omega}{\omega_0}\underline{S}^{\pm} &=& 
\left[\pm 1+\imagu (1-\lambda)\frac{\kappa}{\omega_0}\nabla^2\right]\underline{S}^{\pm}
\mp \frac{1}{2}\left(1-\frac{\tau\chi(1-\lambda)\kappa}{\chi_0}
\nabla^2\right) \left(e^{\mp\imagu\theta} \underline{d}_z\right) ,
 \\
\frac{\omega}{\omega_0}\underline{d}_z &=&
\frac{\Omega^2}{\omega_0^2}(\mathcal{D}+U_\perp)
\left( e^{-\imagu\theta} \underline{S}^{-} - e^{\imagu\theta}
\underline{S}^{+}\right)  -
\imagu\frac{\tau\chi\Omega^2(1-\lambda)}{\chi_0\omega_0\lambda}
(\mathcal{D}+U_\perp) \underline{d}_z.
\end{eqnarray}\end{subequations}
The equation for the power absorption does not need any
modifications. One must be careful since the validity region of the
hydrodynamic approximation is not very large. For temperatures near
$T_{\rm c}$ one must, for typical values of $\tau$, have $B_0 \lesssim
15$ mT. We mainly used the hydrodynamic approximation for checking our
calculations in the limit of $\tau \rightarrow 0$.

For longitudinal case, where ${\bf B}' = B'\hat{\bf z} \parallel {\bf
B}_0$, we also define dimensionless quantities by writing ${\bf
S}'=\hat{\bf z}\lambda_{\rm d} S_z/\Omega$,
$\boldsymbol{\eta}' = \hat{\bf z}\lambda_{\rm d} \eta_z/\Omega$,
and ${\bf d}'= d_\theta \hat{\bf z}\times \hat{\bf d}_0$.  The
equations of motions are
\begin{subequations}\label{e.eigvpargen}
\begin{eqnarray}
\frac{\omega}{\Omega} S_z &=& 
\imagu(1-\lambda)\frac{\kappa}{\Omega} \nabla^2 S_z 
-\imagu\frac{\kappa}{\Omega} \nabla^2 \eta_z
-\imagu(\mathcal{D}+U_\parallel) d_\theta , \\
\frac{\omega}{\Omega} \eta_z &=& 
-\imagu\lambda(1-\lambda)\frac{\kappa}{\Omega}\nabla^2 S_z
-\imagu\left(\frac{1}{\tau\Omega} - \lambda \frac{\kappa}{\Omega}
\nabla^2 \right) \eta_z 
-\imagu (1-\lambda) (\mathcal{D}+U_\parallel) d_\theta , \\
\frac{\omega}{\Omega} d_\theta &=& 
\imagu\left(S_z +\frac{\chi\eta_z}{\lambda\chi_0} \right)
-\imagu\frac{\gamma B'}{\Omega} ,
\end{eqnarray}\end{subequations}
and the homogeneous adjoint problem for this is given by
\begin{subequations}\label{e.eigvpargen2}
\begin{eqnarray}
\frac{\omega}{\Omega} \underline{S}_z &=& 
\imagu(1-\lambda)\frac{\kappa}{\Omega} \nabla^2 \underline{S}_z
-\imagu\lambda(1-\lambda)\frac{\kappa}{\Omega}\nabla^2 \underline{\eta}_z 
+ \imagu \underline{d}_\theta , \\
\frac{\omega}{\Omega} \underline{\eta}_z &=&
-\imagu\frac{\kappa}{\Omega}\nabla^2 \underline{S}_z
-\imagu\left(\frac{1}{\tau\Omega} - \lambda \frac{\kappa}{\Omega}
\nabla^2 \right) \underline{\eta}_z
+\imagu \frac{\chi}{\lambda\chi_0} \underline{d}_\theta , \\
\frac{\omega}{\Omega} \underline{d}_\theta &=& 
-\imagu (\mathcal{D}+U_\parallel) \left[ \underline{S}_z +
(1-\lambda)\underline{\eta}_z \right] .
\end{eqnarray}\end{subequations}
Power absorption for longitudinal case is given by
\begin{eqnarray}
P(\omega) &=& \frac{\chi \Omega \omega}{2\mu_0\gamma}
\sum_j {\rm Im}\left( c_j \int  B'^* S_{z,j} d^3r \right),
\label{e.powerabspar}
\end{eqnarray}
where the $c_j$'s are the coefficients of the full solution 
$\psi = (S_z\ \eta_z \ d_\theta)^T = \sum_j c_j (S_{z,j}\ \eta_{z,j} \
d_{\theta,j})^T $ and given by 
\begin{eqnarray}
c_j &=& \frac{\imagu\gamma B'}{\omega_j-\omega}\int 
\underline{d}_{\theta,j} d^3r .
\end{eqnarray}
Similar to the transverse case one could write the hydrodynamic
equations using only $d_\theta$ and $S_z$.

The numerical solution for these different eigenvalue problems is
obtained by dividing the calculation length $x=-L/2,\ldots, L/2$, for
example, to 1000 points and approximating the spatial derivatives by
differences. The eigenvalues and vectors of the resulting sparse
5000$\times$5000 (or in the longitudinal case 3000$\times$3000)
matrix $A$ are then solved by {\small MATLAB}. We make use of sparse
matrices and calculate normally only $20$ lowest eigenvalues with
${\rm Re}[\omega_k] > 0$ that give the main contribution to the
absorption spectrum. The complex mode frequencies $\omega_k$ are
related to the reduced frequency shifts $\alpha_{k}$ as given by
Eqs.\ (\ref{e.freqll}) and (\ref{e.freqperpgen}).  

\begin{figure}[!tb] 
\begin{center} 
\includegraphics[width=8cm]{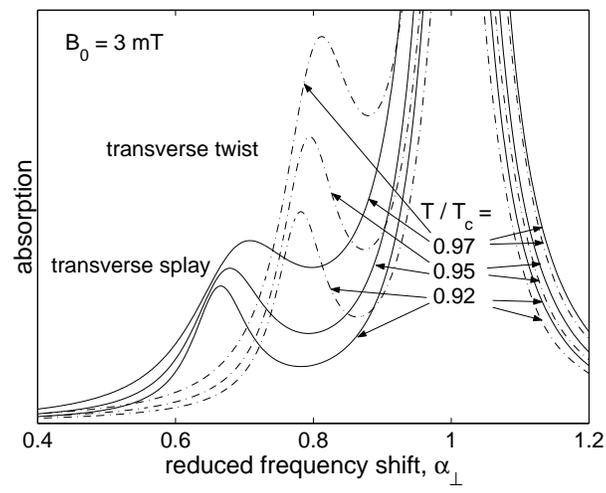}
\caption{Absorption spectra for transverse twist soliton (dash-dotted
lines) and splay soliton (solid lines) at $T/T_{\rm c} = 0.92$,
0.95, and 0.97 when $L=50\xi_{\rm d}$, $B_0 = 3$ mT, 
$\tau=[1.70+ 6.71(1 - T/T_{\rm c})](1+F_0^a)10^{-7}$ s, and 
$\tau_{\rm D} = 1.25\times 10^{-7} T^{-2}$ mK$^2$ s. Other parameters
are obtained using weak coupling and pressure of 3.4 MPa with $F_1^a=-1$
and $F_0^a=-0.746$.}  
\label{spectrum} 
\end{center}
\end{figure}

\begin{figure}[!tb]
\begin{center} 
\includegraphics[width=8cm]{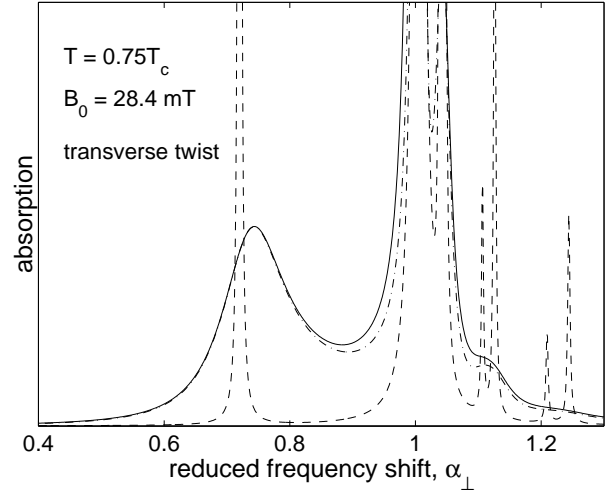}
\caption{Transverse absorption spectrum for a twist soliton
at $T=0.75T_{\rm c}$ and $B_0 = 28.4$ mT (solid line). Other parameters
are the same as in Fig.\ \protect\ref{spectrum}. The effect of the
two dissipation mechanisms is demonstrated by setting $\tau=0$
(dash-dotted line) and $\tau_{\rm D}=0$ (dashed line).}   
\label{spectrumHighH0} 
\end{center}
\end{figure}

\begin{figure}[!tb] 
\begin{center} 
\includegraphics[width=8cm]{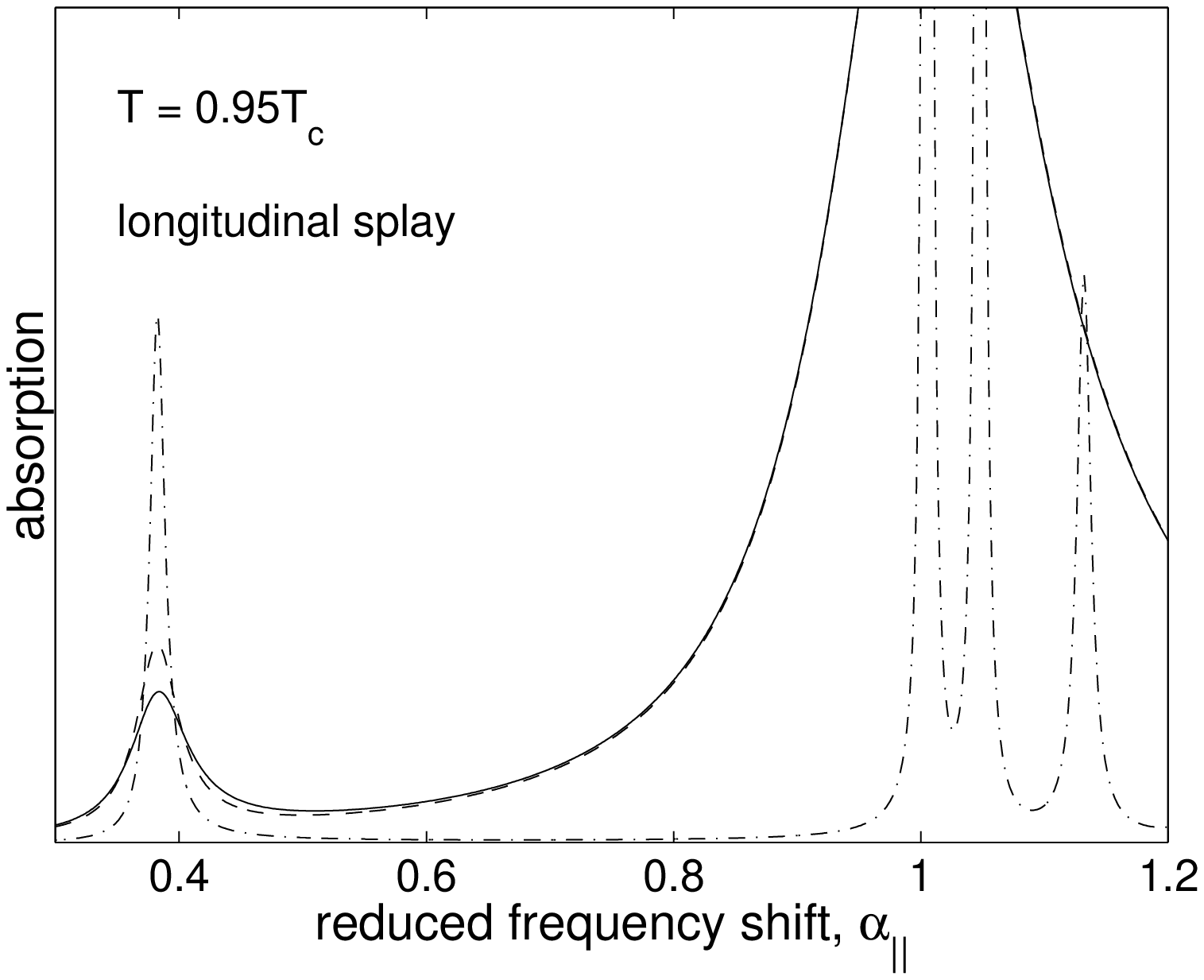}
\caption{Longitudinal absorption spectrum for a splay soliton at
$T=0.95T_{\rm c}$ (solid line). There is no dependence on the magnetic
field, and other parameters are the same as in Fig.\
\protect\ref{spectrum}. The effect of the two dissipation mechanisms is
demonstrated by setting
$\tau=0$ (dash-dotted line) and $\tau_{\rm D}=0$ (dashed line).}   
\label{parallelsplayabs} 
\end{center}
\end{figure}

Additional parameters appear in the calculation compared to the
dissipationless case. The most crucial ones are the relaxation times
$\tau$ and $\tau_{\rm D}$. The Leggett-Takagi time $\tau$ can be
extracted from the width of the bulk peak. Measurements of longitudinal
and transverse resonances are in good agreement.\cite{Gully} We have
reanalyzed the data by Gully {\it et al}.\cite{Gully} including
strong-coupling corrections as described in Sec.\ \ref{s.princnmr} and
get the fit $\tau=[1.70+ 6.71(1 - T/T_{\rm c})](1+F_0^a)10^{-7}$ s in
the range of $T/T_{\rm c}=0.78$--$0.98$. We fix $F_0^a=-0.746$.
For the spin-diffusion coefficient $\kappa$ we use the value $\kappa
T^2 = 1.1\times 10^{-5}$ mK$^2/{\rm s}$ from Refs.
\onlinecite{sachrajda} and \onlinecite{pethick}. This corresponds to
the spin-diffusion time $\tau_{\rm D} = 1.25\times 10^{-7} T^{-2}$ mK$^2$ s.
A few absorption spectra are plotted in Figs.~\ref{spectrum}--\ref{parallelsplayabs}. 

The effect of the normal-superfluid relaxation seems to be simple
line broadening. In the transverse resonance the linewidth measured on
the $\alpha_\perp$ scale is inversely proportional to the magnetic field
$B_0$ for $B_0\gg B_{\rm d}$. This is caused by $\imagu/(\omega_0\tau)$
term in Eq.\ (\ref{e.eigvgenb}).  There
is no field dependence in the longitudinal resonance since
Eqs. (\ref{e.eigvpargen}) do not contain $\omega_0$. In the
longitudinal case the linewidth on the $\alpha$ scale has no strong
temperature dependence and approaches a finite constant when
$T\rightarrow T_{\rm c}$. In the transverse case the linewidth
on the $\alpha$ scale vanishes as $T\rightarrow T_{\rm c}$. All these
characteristics are the same as predicted for the bulk peaks. 

The effect of spin diffusion is more complicated. In addition to
broadening, it shifts the principal soliton peak to higher frequencies.
Since the spin diffusion is due to inhomogeneous texture, its strength
in the transverse case is obtained by comparing the term $\imagu
(1-\lambda)(\kappa/\omega_0) \nabla^2 S^\pm$ with the term
$(\Omega^2/\omega_0^2) \exp(\pm\imagu\theta) (\mathcal{D}+U_\perp)d_z$
in Eq.\ (\ref{e.eigvgena}). Therefore, the effective diffusion constant
$\kappa_{\rm eff}^\perp = (1-\lambda) \kappa \omega_0/\Omega^2$ is
linearly proportional to the magnetic field. As a consequence also the
reduced frequency $\alpha_\perp$ of the satellite peak is field dependent.   
When $B_0=5$ mT the effect of diffusion is already quite large at
$T=0.95T_{\rm c}$ and the soliton satellite peak is almost smeared out.
In the longitudinal case there is no field dependence and the
effective diffusion constant reduces to $\kappa_{\rm eff}^\parallel =
(1-\lambda) \kappa/\Omega$. In both transverse and longitudinal cases the
effective diffusion coefficients $\kappa_{\rm eff}^\parallel$ and 
$\kappa_{\rm eff}^\perp$ diverge when $T\rightarrow T_{\rm c}$, as
$\Omega\rightarrow 0$. 

The relative contributions of normal-superfluid relaxation and spin
diffusion are illustrated in
Figs.\ \ref{spectrumHighH0} and \ref{parallelsplayabs}, where both are
separately shut off by setting $\tau=0$ or $\tau_{\rm D}=0$. It can be
seen that the  broadening of the longitudinal satellite
peak is mostly caused by normal-superfluid conversion  at
$T=0.95 T_{\rm c}$. In the transverse case at high field the spin
diffusion becomes the dominant relaxation mechanism, as well as
approaching $T_{\rm c}$. In uniform order parameter the width of the
bulk peak comes solely from normal-superfluid relaxation. Our soliton
lattice has many higher peaks in the absence of dissipation, but
dissipation seems to remove them. Note that the higher peaks are
mostly suppressed by spin diffusion in the case of Fig.\
\ref{spectrumHighH0} but by normal-superfluid relaxation in the case
of Fig.\ \ref{parallelsplayabs}.

The relatively high field in Fig.\ \ref{spectrumHighH0} has been
chosen to allow a comparison to the spectrum of Hakonen {\it et
al}.\cite{Hakonen} The shape of the twist-soliton spectra are very
similar even if the temperature is quite low, where our modeling of
the spin diffusion is under suspect. Further spectra at lower field
are shown by Parts {\it et al}.\cite{exp} The comparison to Ref.\
\onlinecite{GouldBartolacBozler} is a different case. In this
experiment such a dense soliton lattice was created that no bulk
peak was observable. Avenel {\it et al}.\cite{Avenel} and Gould and
Lee\cite{GouldLee} show soliton spectra measured by sweeping
temperature.   In all cases a proper comparison to our calculations
would require first estimation of the density of solitons and then
correcting our calculations for that density, which we have not done.  
What we can state that there seems to be no obvious contradictions
between theory and experiment concerning the linewidth.  

\begin{figure}[!tb]
\begin{center} 
\includegraphics[width=8cm]{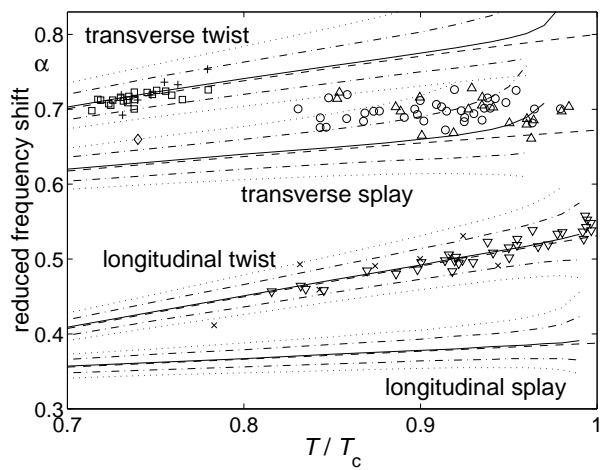}
\caption{The resonance frequencies and linewidths as a function of
temperature at $B_0=3$ mT. The solid lines give the frequency of
maximum  satellite absorption, the dash-dotted lines 75\%,  and dotted
lines  50\% of the maximum. The dashed lines give resonance frequency
in the absence of dissipation. Other parameters are $L = 50\xi_{\rm d}$, 
$\tau=[1.70+ 6.71(1 - T/T_{\rm c})](1+F_0^a)10^{-7}$ s, 
$\tau_{\rm D} = 1.25\times 10^{-7} T^{-2}$ mK$^2$ s,
pressure 3.4 MPa  ($F_1^s=14.45$, $F_1^a=-1$, and $F_0^a=-0.746$ and
weak coupling). The experimental points are the same as in Fig.
\ref{R2}.} 
\label{r2spindiff} 
\end{center}
\end{figure}

A summary of our results at $B_0=3$ mT is shown Fig.\ \ref{r2spindiff}.
We see that the frequency of maximum absorption is shifted towards
higher frequencies, especially at temperatures near $T_{\rm c}$. This
alleviates the disagreement found between dissipationless
theory and the experiment by Gould and
Lee\cite{GouldLee} in the transverse mode of splay soliton.
However, the experimental data does not show any sign of the divergence
at $T_{\rm c}$ predicted by our model. 
The shift of the resonance frequency results from spin diffusion. For
the longitudinal mode it is field independent, but for the transverse
mode it increases with increasing field. In high field this shift can
be substantial even at $T\approx 0.7T_{\rm c}$, see Fig.\
\ref{spectrumHighH0}. This poses a problem since no
field-dependent shift has been observed experimentally. This is the
case particularly for the transverse mode of twist
soliton measured at $B_0=9.9$ mT
($\square$) (Ref. \onlinecite{exp}) and 28 mT ($+$) (Ref. \onlinecite{Hakonen}).
Similar evidence for transverse mode of splay soliton measured   
at $2.5 \ldots 3.7$ mT ($\circ$) (Ref. \onlinecite{GouldLee}) and at $15\ldots20$
mT ($\triangle$) (Ref. \onlinecite{GouldBartolacBozler}) is not so
clear since the latter data is measured under different conditions, as
discussed above.

\section{CONCLUSIONS}

The frequencies of soliton satellite peaks in the NMR spectrum are
calculated at all temperatures. The agreement of dissipationless
theory with experiments is very good. However, there is a small
difference in the transverse-mode frequency of a splay soliton near
$T_{\rm c}$. This difference is partly explained by taking into
account spin diffusion. The spin diffusion shifts up
the reduced frequencies $\alpha$ at high fields, which has not
been observed experimentally. We hope that new experiments
could clarify this problem. We also point out the relatively narrow
longitudinal line of the splay soliton, which has not been studied
experimentally. 

It would be of interest to extend the present calculations to vortex
lines in $^3$He-A. This requires two-dimensional calculation and thus
would be computationally much more demanding than the present one.

\section*{ACKNOWLEDGMENTS}
We thank Pete Sivonen for contribution in early stages of this work.
This research was supported by the Academy of Finland and the
V\"ais\"al\"a foundation.

\end{document}